\documentstyle[11pt]{article}

\newcommand{\bfp}{\mbox{\bf p}}
\newcommand{\bfu}{\mbox{\bf u}}
\newcommand{\bfe}{\mbox{\bf e}}

\begin{document}

\vspace*{1.in}
\begin{center}

{\Large {\bf  Gravity, Geometry, and Equivalence }} \\

\vspace{.5in}

{\large {\bf  Kenneth Dalton }} \\
Post Office Box 587 \\
Mammoth Hot Springs \\
Yellowstone Park WY 82190, U.S.A. \\
\vspace{1.5in}

{\bf Abstract}

\end{center}

\vspace{.25in}

We show that the energy-momentum four-vector of a planet,
$ {\bfp} = m {\bfu} $, is conserved during
geodesic motion.  Therefore, there is no exchange of energy-
momentum with the gravitational field.  We discuss the
meaning of a gravitational field which is free of energy,
momentum, and stress.

\clearpage

The conclusions to be drawn in this paper all derive from the
following elementary calculation.  We consider a planet or
satellite in geodesic motion and treat it as a point
particle, with energy-momentum four-vector (c = 1)

\begin{equation}
   {\bfp} = m {\bfu}
\end{equation}
We now inquire as to the rate of change of this four-vector.
Expanding the velocity
$ {\bfu} = {\bfe}_\mu u^\mu $
and making use of Cartan's derivative operator [1,2]

\begin{equation}
   d {\bfe}_\mu = {\bfe}_\lambda \Gamma^\lambda_{\mu\nu} \, dx^\nu
\end{equation}
we form the expression

\begin{eqnarray}
   \frac{d {\bfp}}{ds} = m \frac{d {\bfu}}{ds}
   & = & m \left\{ {\bfe}_\mu \frac{du^\mu}{ds}
       + \frac{d {\bfe}_\mu}{ds} u^\mu \right\}  \nonumber \\
   & = & m {\bfe}_\mu \left\{ \frac{du^\mu}{ds}
       + \Gamma^\mu_{\nu\lambda} u^\nu u^\lambda \right\}
\end{eqnarray}
This quantity is zero along any geodesic path.  Therefore,
the energy-momentum of the planet is conserved

\begin{equation}
   \frac{d{\bfp}}{ds} = 0
\end{equation}
From this we conclude that there is no exchange of energy-
momentum with the gravitational field; in other words, there
is no gravitational force acting on the planet [3,4].

In his earlier work, Einstein identified the term
$ \Gamma^\mu_{\nu\lambda} u^\nu u^\lambda $
with just such a gravitational force.  For example, he states that this
term is ``an expression for momentum, and for energy, as transferred per
unit of volume and time from the gravitational field to matter'' [5].
However, it is known that all connection coefficients
$ \Gamma^\mu_{\nu\lambda} $  can be transformed to zero in any given
infinitesimal region (geodesic coordinates) [6].  Moreover, it has been
shown that the $ \Gamma^\mu_{\nu\lambda} $  can be transformed to zero
not just at one point but at all points along any given world-line [6,7].
This applies, in particular, to all points along any given planetary
orbit. We  conclude that no force of any kind may be
associated with geodesic motion.
   \footnote{\small It is sometimes stated that the transformation
$ \Gamma^\mu_{\nu\lambda}  \longrightarrow  0 $
eliminates the gravitational field from the region in question.
However, the metric $ g_{\mu\nu} $ , the curvature
$ R^\mu{}_{\nu\lambda\rho} $ , and the geodesic lines
all represent invariant geometric properties of the underlying
space-time, which are not altered by a change of coordinates:
the geodesics are not `straightened' nor is the space-time
region `flattened' by this procedure.}

The absence of any transfer of energy-momentum during
planetary motion shows that the gravitational field
has no energy, momentum, or stress.
The question, therefore, is this:
What is the meaning of an
energy-free gravitational field?  For the answer, we turn to
Einstein's principle of equivalence.  In his celebrated
``elevator experiment,'' he showed that gravity is equivalent
to a kinematical effect, namely, acceleration.  Then, in
deducing the red-shift of light, he argued that clocks,
located at points of differing gravitational potential, must
run at different rates---again, a kinematical effect.
Finally, he described planetary motion by means of the
geodesic equation, that is, without the introduction of
inertial mass.  To this we add the energy-free gravitational
field.  It forms a curved ``kinematical background'' for
planetary motion, much as Newtonian space and time once
formed a flat kinematical background.  This complete
reduction to kinematics yields a purely geometric field of
gravitation (composed of metric, curvature, geodesic lines)
and a theory which is fully consistent with the original
guiding principle of equivalence.

\section*{\large {\bf References}}

\begin{enumerate}

\item A. Lichnerowicz, {\it Elements of Tensor Calculus}
      (John Wiley, New York, 1962) chapter V.
\item J. Vargas and D. Torr, {\it Gen.Rel.Grav.} {\bf 23}, 713
      (1991).
\item K. Dalton, {\it Gen.Rel.Grav.} {\bf 21}, 533 (1989).
\item K. Dalton, {\it Hadronic J.} {\bf 17}, 139 (1994).
\item A. Einstein, ``The Foundation of the General Theory of
      Relativity'' in {\it The Principle of Relativity}
      (Dover, New York, 1952) page 151.
\item L. Landau and E. Lifshitz, {\it The Classical Theory of
      Fields} (Pergamon, Elmsford, 4th ed., 1975) page 241.
\item T. Levi-Civita, {\it The Absolute Differential
      Calculus} (Dover, New York, 1977) page 167.

\end{enumerate}

\end{document}